\documentclass[11pt,a4paper]{article}

\usepackage[a4paper,margin=2.7cm]{geometry}
\usepackage{amsmath,amssymb,amsfonts}
\usepackage{physics}
\usepackage{bm}
\usepackage{booktabs}
\usepackage{xcolor}
\usepackage{hyperref}
\usepackage{pgfplots}
\pgfplotsset{compat=1.18}

\hypersetup{
  colorlinks=true,
  linkcolor=blue,
  citecolor=blue,
  urlcolor=blue
}
\newcommand{\Msun}{\ensuremath{M_\odot}}
\newcommand{\Mearth}{\ensuremath{M_\oplus}}
\newcommand{\Mjup}{\ensuremath{M_J}}
\newcommand{\Mnep}{\ensuremath{M_{\rm Nep}}}
\newcommand{\Msat}{\ensuremath{M_{\rm Sat}}}
\newcommand{\pc}{\ensuremath{\mathrm{pc}}}
\newcommand{\AU}{\ensuremath{\mathrm{AU}}}
\newcommand{\rhoDM}{\ensuremath{\rho_{\rm DM}^{\rm loc}}}

\title{A Solar-System Window for Hidden Stellar Companions}

\author{
Karim Benakli\\
{\small Sorbonne Universit\'e, CNRS,}\\{\small   Laboratoire de Physique Th\'eorique et Hautes \'Energies, LPTHE,}\\{\small  F-75005 Paris, France}\\
{\small \texttt{kbenakli@lpthe.jussieu.fr}}
}

\date{ }

\begin{document}
\maketitle

\begin{abstract}
Could the closest stellar or substellar object to the Sun be not an ordinary star at
parsec distance, but a hidden-brane companion at hundreds or thousands of astronomical
units? We do not perform a new Solar-System dynamics analysis; instead we construct a
phenomenological mass-distance map using an illustrative ephemeris tidal envelope
calibrated to Planet-Nine-like constraints. A smooth dark matter halo cannot supply such
an object: the local density contains only a sub-Pluto mass within \(1000\,\AU\). A nearby
hidden-brane companion must therefore be a structured, gravitationally bound object rather
than a typical halo draw. The illustrative envelope still allows Earth to sub-Saturn
masses at \(300\text{--}1000\,\AU\), rising to Jovian mass near \(2000\,\AU\). In a simple
QCD-scaled hidden sector with confinement scale larger than the ordinary one by a factor
\(\kappa\sim10\), the minimum hidden stellar mass overlaps the upper part of this window,
providing a benchmark for a genuine hidden-sector star: bright in dark photons,
electromagnetically dark to us, and visible only through gravity. We also derive an
Earth-based source-strength proxy for brane-to-brane channels and show that it grows as
\(d^2\) along the ephemeris envelope: the largest received source scale comes from the most
massive companion still allowed, not the nearest one. A probe sent to the companion's
gravitational projection would reduce the ordinary source--receiver separation by two to
three orders of magnitude relative to Earth-based operation. This is not a detection
forecast; the excitation of KK modes, the compact-direction brane-to-brane transfer factor,
and the detector response remain model-dependent.
\end{abstract}



\section{Introduction}

The nearest known star is Proxima Centauri, at a distance of about 
\[
  1.3\,\pc \simeq 2.7\times10^5\,\AU .
\]
We say this so routinely that it is easy to lose sight of how much it assumes. The
statement is about ordinary stars, made of ordinary matter, radiating ordinary photons.
The question pursued in this note is whether a different kind of star---one whose light is
not our light---could be much closer. Not a cold asteroid, and not merely a dark planet,
but an active astrophysical object, sitting somewhere between the orbits of the outer
planets and the inner Oort cloud, and invisible to ordinary electromagnetic surveys.

The candidate we have in mind is a hidden-sector star: an object made of matter that does
not couple, or couples only negligibly, to the electromagnetic field on our brane. Such an
object could shine in its own hidden photons, cool and burn in its own sector, and remain
undetectable in optical, infrared, and radio surveys. Gaia, WISE, 2MASS, and related surveys~\cite{Gaia:2016,WISE:2010,TwoMASS:2006}
would not identify it as a faint star, because they rely on visible-sector photons. 
Gaia-like astrometry or microlensing could still constrain it gravitationally, but only
through the mass it projects onto our brane. Its presence would therefore have to be
inferred from gravity: planetary perturbations, trans-Neptunian dynamics, long-period
comets, astrometric lensing, or eventually spacecraft tracking.

Brane-world scenarios make this idea concrete in a broad sense. Matter localized on a
hidden brane is electromagnetically invisible on ours, while gravity propagates through
the higher-dimensional bulk and couples to both sectors. The details depend on the
underlying compactification and on the mechanism localizing matter on the two branes; we
will not need them here. The only assumptions used below are modest: hidden matter is
localized away from our brane, the ordinary long-distance force is dominated by the
massless four-dimensional graviton, and the compactification or brane-separation scale is
microscopic compared with the astronomical distances of interest. In the benchmark used
below, this scale is sub-micrometric. On Solar-System scales, a compact object on a hidden
brane therefore gravitates, to excellent approximation, like an ordinary compact mass.

The natural question is then unavoidable. Could the closest stellar object to the Sun be
not an ordinary star at parsec distance, but a hidden-brane star, or a hidden stellar
remnant, at hundreds or thousands of astronomical units?

There are two sharply different ways such an object might be imagined, and they lead to
very different constraints. The first is as a member of the smooth Galactic dark matter
halo, a MACHO-like object that happens to lie close to the Sun. For planetary or stellar
masses this picture is not viable. The local dark matter density is simply too small:
there is not enough mass within \(10^3\text{--}10^4\,\AU\) for an Earth-, Neptune-,
Jovian-, or stellar-mass companion to arise as a typical draw from a smooth distribution.
This conclusion is reached before any microlensing or compact-object abundance bound is
applied. Such bounds can be folded in on top, and they reinforce the result, but the local
mass budget is already decisive.

The second possibility is qualitatively different, and it is the one this paper examines.
If the hidden sector is structured and dissipative---capable of cooling, fragmenting, and
assembling bound objects, much as ordinary baryons did before forming stars---then a nearby
hidden object is not a Poisson draw from the smooth halo. It is a bound companion of the
Solar System, formed alongside it or captured later. The relevant constraints are then not
primarily those of Galactic halo statistics, but local Solar-System constraints:
planetary ephemerides, ranging data, trans-Neptunian dynamics, long-period comet
dynamics, astrometric lensing, and, in principle, direct in situ spacecraft tracking.

This distinction is worth setting against a related body of work that has already asked
whether the putative Planet Nine~\cite{Batygin:2019} might be an exotic compact object
rather than an ordinary planet. Scholtz and Unwin proposed that a Planet-Nine-mass body in
the outer Solar System could be a primordial black hole, with a capture probability
comparable to that of a free-floating planet~\cite{Scholtz:2019csj}; subsequent work
explored the electromagnetic signatures that accretion onto such a black hole might leave
behind~\cite{Siraj:2020upy}. Witten took a different approach, proposing that a primordial
black hole or other exotic compact object at the location of Planet Nine could instead be
searched for directly, by sending a fleet of small, fast spacecraft to probe its gravitational
field~\cite{Witten:2020qew}. That local-probe idea reappears later in this paper, but in a
different role: not only as a way of mapping the companion's gravity, but as a way of
shortening the ordinary source--receiver distance in a possible brane-to-brane channel.

The scenario considered here is different in both origin and phenomenology. The object is
not a primordial black hole but a structured hidden-sector companion, possibly a genuine
hidden star or substellar object, whose luminosity is carried by hidden photons rather
than by ordinary accretion-powered emission. Its gravitational constraints overlap with
the Planet-Nine mass-distance window, but its microphysics and possible brane-to-brane
signals are distinct.

This question is also motivated by brane-to-brane gravitational communication. If two
branes can exchange gravitational signals through Kaluza--Klein modes in the bulk, as in
the channel considered in the companion paper~\cite{Benakli:2026qag}, then a
hidden-brane object close to the Sun would not merely be an exotic astronomical
possibility. It would be the most favourable nearby source or target for such a channel. A
companion at \(10^3\text{--}10^4\,\AU\) would replace the nearest ordinary star, at
\(2.7\times10^5\,\AU\), by an object one to three orders of magnitude closer, with the
corresponding geometric advantage. Whether such a companion is allowed by existing
Solar-System observations is therefore a prerequisite question.

This is not a new Solar-System dynamics analysis. It is a phenomenological map of the
mass-distance window in which a hidden stellar companion could evade existing local
constraints, together with a hidden-sector benchmark showing that stellar masses in this
window are not parametrically absurd. The purpose of this note is therefore to make the
question quantitative without pretending to replace a full ephemeris fit.

We first show, in Section~\ref{sec:smooth_halo}, that the smooth-halo picture cannot
supply an astrophysically interesting hidden object at Solar-System distances, by a wide
margin and for density-budget reasons alone. We then turn, in Section~\ref{sec:bound_companion},
to the bound-companion alternative and estimate the mass-distance window left open by
Solar-System dynamics, using an illustrative tidal envelope calibrated against
Planet-Nine-like constraints. Within that window, Section~\ref{sec:population_dark_sector}
identifies a simple QCD-scaled hidden-sector benchmark in which the minimum hidden stellar
mass is pushed down into the Jovian range. Section~\ref{sec:source_strength_proxy} then
introduces a model-independent source-strength proxy for any brane-to-brane channel,
explains why a static or slowly orbiting companion cannot by itself excite the massive KK
tower, and identifies the kind of microscopic time dependence that would be required.
Finally, Section~\ref{sec:local_transmission} points out that, if such a channel exists,
its geometry strongly favours operation near the companion rather than near the Earth,
with amplitude gains of \(10^2\text{--}10^3\) and flux gains of \(10^4\text{--}10^6\) in
principle.

None of this is a detection claim. The aim is to identify a dynamically allowed corner of
parameter space in which a hidden stellar companion of the Sun could exist, to exhibit a
simple hidden-sector benchmark that naturally places an object in that window, and to
clarify how the geometry of a brane-to-brane channel would favour any future local test.



\section{The smooth-halo picture and its limits}
\label{sec:smooth_halo}

Suppose hidden-brane compact objects simply trace the smooth Galactic dark matter halo, the
way ordinary MACHOs would. Their local abundance is then fixed by nothing more than the
local dark matter density and their mass. Whether the objects in question are stars,
compact remnants, planets, or smaller bodies makes no difference to this argument: in a
smooth halo, the local number density follows directly from the available local mass density. We
work through this picture in three steps. First, for a population of a single mass, we ask
how far away the nearest object is expected to be. We then turn the question around and ask
what mass a smooth halo can plausibly deliver within a given distance. Finally, we connect
this density argument to the more familiar MACHO and compact-dark-matter constraints, and
show that they only sharpen the conclusion.

\subsection{Nearest-neighbour distance}

Let \(\rhoDM\) denote the local dark matter density. We take as a conservative reference
value~\cite{Read:2014qva,deSalas:2020hbh}
\begin{equation}
  \rhoDM \simeq 0.008\,\Msun/\pc^3
  \simeq 0.3\,{\rm GeV}/{\rm cm}^3 .
  \label{eq:rhoDM}
\end{equation}
This is consistent with standard local determinations~\cite{Bovy:2012tw}; some recent
Galactic mass models prefer somewhat larger values, closer to
\(0.4\text{--}0.5\,{\rm GeV}/{\rm cm}^3\). Our choice is conservative in the relevant
sense: a larger \(\rhoDM\) would increase the smooth-halo mass available within a given
Solar-System volume, or equivalently shrink the nearest-neighbour distance for objects of
fixed mass. Even such larger values would only strengthen the smooth-halo supply by
factors of order unity, and do not change the conclusion below.

If a fraction \(f_B\) of this density is carried by hidden-brane compact objects of a
common mass \(M\), their local number density is simply
\begin{equation}
  n_B(M)
  =
  \frac{f_B\rhoDM}{M},
  \label{eq:nB}
\end{equation}
and the expected number of such objects inside a sphere of radius \(R\) around the Sun is
\begin{equation}
  N_B(<R)
  =
  \frac{4\pi}{3}R^3 n_B
  =
  \frac{4\pi}{3}R^3\frac{f_B\rhoDM}{M},
  \label{eq:NB}
\end{equation}
or numerically,
\begin{equation}
  N_B(<R)
  \simeq
  33.5\,f_B
  \left(\frac{R}{10\,\pc}\right)^3
  \left(\frac{M}{\Msun}\right)^{-1}.
  \label{eq:NBnum}
\end{equation}
For a homogeneous Poisson process, the probability of finding no object at all within
radius \(r\) is
\begin{equation}
  P(>r)
  =
  \exp\!\left[
  -\frac{4\pi}{3}n_B r^3
  \right],
  \label{eq:poisson}
\end{equation}
and setting \(P(>r_{\rm med})=1/2\) gives the median nearest-neighbour distance,
\begin{equation}
  r_{\rm med}
  =
  \left(
  \frac{3\ln2}{4\pi n_B}
  \right)^{1/3}
  =
  \left(
  \frac{3\ln2}{4\pi}
  \frac{M}{f_B\rhoDM}
  \right)^{1/3}.
  \label{eq:rmed}
\end{equation}
For solar-mass hidden objects accounting for all of the local dark matter
(\(M=\Msun\), \(f_B=1\)), this gives
\begin{equation}
  r_{\rm med}
  \simeq
  2.7\,\pc
  \simeq
  5.6\times10^5\,\AU.
  \label{eq:rmed_solar}
\end{equation}
The nearest such object would sit at the same kind of distance as the nearest ordinary
stars, several light-years away. Long before any microlensing survey or dynamical bound
enters the picture, the sheer thinness of the local dark matter density already keeps a
smooth halo of solar-mass hidden objects well clear of the Solar System.

\subsection{The mass scale selected by a given distance}

Equation~\eqref{eq:rmed} scales as \(M^{1/3}\): lighter objects are more numerous and
therefore sit closer. It is natural to ask the question the other way round. Rather than
fixing the mass and asking how far the nearest object is, fix the distance and ask which
mass the halo could plausibly supply there.

Inverting Eq.~\eqref{eq:rmed} gives
\begin{equation}
  M_{\rm med}(d)
  =
  \frac{4\pi}{3\ln2}
  f_B\rhoDM\, d^3.
  \label{eq:Mmed}
\end{equation}
This is not the total dark matter mass enclosed within radius \(d\) — that quantity would
be \((4\pi/3)\rhoDM d^3\), without the \(\ln 2\) factor. \(M_{\rm med}(d)\) answers a
different and more useful question: for which monochromatic object mass is the median
nearest-neighbour distance exactly \(d\)? Above that mass, the nearest representative
typically lies farther out than \(d\); below it, typically closer.

For \(f_B=1\) and the reference density of Eq.~\eqref{eq:rhoDM},
\begin{equation}
  M_{\rm med}(d)
  \simeq
  5.5\times10^{-9}\,\Msun
  \left(\frac{d}{1000\,\AU}\right)^3.
  \label{eq:Mmed_num}
\end{equation}
At \(d=1000\,\AU\) this comes to
\begin{equation}
  M_{\rm med}(1000\,\AU)
  \simeq
  1.8\times10^{-3}\,\Mearth,
  \label{eq:Mmed_1000}
\end{equation}
just under the mass of Pluto, \(M_{\rm Pluto}\simeq2.2\times10^{-3}\,\Mearth\). Push out to
\(d=2000\,\AU\) and the budget only climbs to
\begin{equation}
  M_{\rm med}(2000\,\AU)
  \simeq
  4.4\times10^{-8}\,\Msun
  \simeq
  4.6\times10^{-5}\,\Mjup,
  \label{eq:Mmed_2000}
\end{equation}
and even at \(d=10^4\,\AU\), deep into the Oort cloud,
\begin{equation}
  M_{\rm med}(10^4\,\AU)
  \simeq
  5.5\times10^{-6}\,\Msun
  \simeq
  1.8\,\Mearth.
  \label{eq:Mmed_10000}
\end{equation}
The pattern that matters for the rest of this paper is unambiguous. A hidden-brane object
of Neptune, Jovian, or stellar mass anywhere between \(10^3\) and \(10^4\,\AU\) cannot be a
typical member of a smooth local halo, even granting the maximally generous assumption that
all of the local dark matter density is carried by such objects. Even an Earth-mass object is
already excluded as a typical smooth-halo draw at \(10^3\,\AU\), and only becomes a
borderline case once one reaches distances approaching \(10^4\,\AU\). Lowering the
compact-object fraction \(f_B\) cannot rescue the picture, since it only pushes the nearest
neighbour farther away, as \(r_{\rm med}\propto f_B^{-1/3}\). We return to this comparison
explicitly in Fig.~\ref{fig:budget_ephemeris}, once the ephemeris envelope has been
introduced.

\subsection{Relation to MACHO and compact-dark-matter constraints}
\label{subsec:macho}

Everything in the preceding subsection is a statement about local density alone. No
microlensing survey, no wide-binary disruption argument, no stellar-cluster heating bound
has been used. It is worth pausing to see how those more familiar constraints fit alongside
this purely geometric one, and to check that they do not weaken it.

Gravitational microlensing surveys constrain the fraction of the smooth Galactic halo that
can be made of compact objects, as a function of mass
\cite{Tisserand:2006zx,Wyrzykowski:2011tr}. The exclusion curve is far from flat, and the
precise limit depends on \(M\), on the survey, and on the assumed halo model. Global
compilations combining microlensing with other compact-object constraints show that, over
large parts of the planetary-to-stellar mass range relevant here, the allowed smooth-halo
fraction is well below unity, and can fall to the percent or sub-percent level
\cite{Carr:2020gox,Green:2020jor}. For orientation, we will use
\begin{equation}
  f_B(M)
  \lesssim
  3\times10^{-3}
  \label{eq:macho_permille_bound}
\end{equation}
as a representative strong bound in the microlensing-sensitive regime, not as a universal
mass-independent limit. The important point for the present argument is weaker and more
robust: any bound \(f_B<1\) only reduces the smooth-halo mass budget relative to the
maximally generous estimate used above.

The one robustly open region where compact objects can still make up all of the dark
matter in a smooth, monochromatic PBH-like scenario lies at much smaller masses,
\begin{equation}
  M\lesssim 3\times10^{-12}\,\Msun
  \simeq
  10^{-6}\,\Mearth,
  \label{eq:macho_low_mass_window}
\end{equation}
within the assumptions and caveats of the standard PBH constraint compilations
\cite{Carr:2020gox,Green:2020jor}. This is asteroid territory, and of no help for a hidden
stellar or planetary companion.

Folding these bounds in only reinforces the density-budget conclusion, since
\(M_{\rm med}\) scales linearly with \(f_B\):
\begin{equation}
  M_{\rm med}(d;f_B)
  =
  f_B\,M_{\rm med}(d;f_B=1).
\end{equation}
With \(f_B\lesssim3\times10^{-3}\), the budget at \(d=2000\,\AU\) falls to
\begin{equation}
  M_{\rm med}(2000\,\AU;f_B)
  \lesssim
  1.3\times10^{-10}\,\Msun
  \simeq
  1.4\times10^{-7}\,\Mjup,
  \label{eq:Mmed_2000_permille}
\end{equation}
or, read the other way, the nearest-neighbour distance for any fixed mass grows by a factor
\(f_B^{-1/3}\simeq6.9\) once such a compact-object abundance bound is imposed. The
observational limits make a nearby smooth-halo planetary or stellar object even less
plausible than the already-generous \(f_B=1\) estimate suggested.

One caveat deserves a closer look. The bounds just quoted assume a monochromatic mass
function, and a population spread over a range of masses cannot simply inherit them point
by point. Writing \(\psi(M)\equiv df_B/d\ln M\) for the fraction of dark matter in compact
objects per logarithmic mass interval, the local number density becomes
\begin{equation}
  n_B
  =
  \rhoDM
  \int
  \frac{\psi(M)}{M}\,d\ln M,
  \label{eq:n_extended}
\end{equation}
with the enclosed count following as
\begin{equation}
  N_B(<R)
  =
  \frac{4\pi}{3}R^3\rhoDM
  \int
  \frac{\psi(M)}{M}\,d\ln M.
  \label{eq:N_extended}
\end{equation}
Comparing this honestly to microlensing and dynamical constraints means folding the full
mass function against each survey's mass-dependent response, not just reading off a single
number; see, for example, Refs.~\cite{Carr:2020gox,Green:2020jor,Green:2016xgy,Carr:2017jsz}.
In the common approximation where the response is linear in the compact-object fraction, a
useful guide is
\begin{equation}
  \int
  \frac{\psi(M)}{f_{\rm max}(M)}
  \,d\ln M
  \lesssim
  1,
  \label{eq:extended_constraint}
\end{equation}
where \(f_{\rm max}(M)\) is the monochromatic upper bound. This expression is only a guide,
not a substitute for a dedicated likelihood analysis. Spreading the mass function across
several bins can soften sharply localized monochromatic limits, but the relief comes with a
price: a distribution centred in an apparently open window can still have tails leaking
into neighbouring constrained territory, with heavy tails running back into microlensing,
wide-binary, or dynamical-heating bounds. If the compact objects are primordial black
holes, very light tails are also subject to PBH-specific evaporation and survival
constraints. These evaporation bounds, however, do not apply to the hidden-sector stellar
or substellar companions considered in the rest of this paper.

None of this changes the logic of the present paper, however. Extended mass functions and
clustering can modify how halo-wide compact-object bounds should be read, but they cannot
increase the local smooth-halo mass density. If the Sun happened to sit inside an unusually
dense clump of hidden compact objects, the smooth-halo assumption would already have broken
down---that would be a structured local environment, which is precisely the alternative
picture developed in the rest of this paper, not a loophole within the smooth-halo scenario
itself.

The upshot is the same whichever way one approaches it. A Neptune-, Jovian-, or
stellar-mass hidden object at \(10^3\text{--}10^4\,\AU\) should not be regarded as a
typical smooth-halo draw. If such an object exists, it must belong to a structured local
environment, for instance as a gravitationally bound companion of the Solar System. This is
the possibility we turn to next.


\section{Hidden-brane bound companion}
\label{sec:bound_companion}

The previous section ruled out a nearby planetary or stellar hidden object as a typical
draw from a smooth Galactic halo. This leaves a different possibility open: the hidden
sector may be structured and dissipative, capable of forming objects gravitationally bound
to the Solar System rather than distributed as part of the smooth Galactic halo. Such an
object would not represent the local smooth-halo density at all. It would be a local
overdensity, much as an ordinary bound planet or companion is, only localized on the hidden
brane.

Three assumptions underlie what follows. The hidden companion sources the ordinary
long-distance gravitational field through the four-dimensional graviton, so its effect on
planetary dynamics is indistinguishable from that of an ordinary compact mass \(M_X\) at
the same three-dimensional position. It emits no visible photons on our brane: it may be
bright in hidden photons, but these do not couple appreciably to our electromagnetic
sector. And we do not assume that such objects make up any significant fraction of the
smooth Galactic dark matter halo. The constraints considered here are therefore
Solar-System dynamical constraints on a single bound companion, not MACHO abundance
constraints on a halo population---a distinction worth making precise before going
further.

The relevant constraints are purely gravitational:
\begin{itemize}
  \item planetary ephemerides,
  \item Cassini and other ranging data,
  \item perturbations of trans-Neptunian objects,
  \item long-period comet dynamics,
  \item Solar-System barycentre acceleration,
  \item stellar encounters and Galactic tides at very large distances.
\end{itemize}

\subsection{Why standard MACHO microlensing bounds do not directly apply}

The MACHO limits discussed in Section~\ref{subsec:macho} constrain a statistical population
of compact lenses spread through the Galactic halo
\cite{Tisserand:2006zx,Wyrzykowski:2011tr,Carr:2020gox,Green:2020jor}. They say nothing
directly about a single Solar-System companion. A bound hidden object at
\(d\sim10^3\text{--}10^4\,\AU\) occupies one fixed sky position and follows one specific
orbit; it does not sample the lines of sight toward the Magellanic Clouds, the Galactic
bulge, or Andromeda in the way a smooth halo population does.

This does not mean such a companion is invisible to lensing in principle. If it gravitates
on our brane, it can deflect the light of background stars or shift their apparent
positions, just as any compact mass would. For a lens at distance \(d\) and a background
source at \(D_S\gg d\), the Einstein angle is
\begin{equation}
  \theta_E
  =
  \left[
  \frac{4GM_X}{c^2}
  \frac{D_S-d}{dD_S}
  \right]^{1/2}
  \simeq
  \left(
  \frac{4GM_X}{c^2 d}
  \right)^{1/2},
  \label{eq:thetaE_local}
\end{equation}
with a corresponding physical Einstein radius in the lens plane,
\begin{equation}
  R_E=d\theta_E
  \simeq
  \left(
  \frac{4GM_X d}{c^2}
  \right)^{1/2}.
  \label{eq:RE_local}
\end{equation}
For a Jovian-mass companion at \(d=2000\,\AU\), this gives
\begin{equation}
  R_E
  \simeq
  4.1\times10^7\,{\rm m}\,
  \left(\frac{M_X}{\Mjup}\right)^{1/2}
  \left(\frac{d}{2000\,\AU}\right)^{1/2},
  \label{eq:RE_Jupiter}
\end{equation}
or equivalently
\begin{equation}
  \theta_E
  \simeq
  28\,{\rm mas}\,
  \left(\frac{M_X}{\Mjup}\right)^{1/2}
  \left(\frac{2000\,\AU}{d}\right)^{1/2}.
  \label{eq:thetaE_Jupiter}
\end{equation}

A few features of this estimate are worth pointing out. Since the companion sits on a
hidden brane, it is not an opaque body occupying space on our own; it need not occult the
lensed light at all, even though it may have a finite physical radius in its own sector.
Finite-size effects can still matter gravitationally if the mass distribution turns out to
be extended compared with \(R_E\), but the familiar worry about an ordinary planet
physically blocking the lensed images simply does not arise here. The apparent motion of
such a nearby object is also dominated by annual parallax, an effect that is subdominant
for the distant halo lenses targeted by standard microlensing surveys. At \(d=2000\,\AU\),
the parallax amplitude is
\begin{equation}
  \pi\simeq \frac{1\,\AU}{d}\simeq 100\,{\rm arcsec}.
\end{equation}
Thus any lensing or astrometric signal would be strongly modulated by the Earth's own
motion and would differ qualitatively from a standard halo microlensing event. The
associated Einstein-crossing time, set by the parallactic motion rather than by the lens's
own orbital speed, is of order
\begin{equation}
  t_E
  \sim
  20\,{\rm min}\,
  \left(\frac{M_X}{\Mjup}\right)^{1/2}
  \left(\frac{d}{2000\,\AU}\right)^{1/2},
  \label{eq:tE_local}
\end{equation}
up to geometric factors that depend on the companion's position on the sky. A bound hidden
companion would therefore call for targeted astrometric lensing searches,
occultation-like surveys if any visible-sector coupling exists at all, or precision
dynamical reconstruction. We do not pursue any of these here; the rest of this section
relies on the most robust and model-independent handle available, which is Solar-System
dynamics itself.



\subsection{Ephemeris tidal envelope}

For a companion at heliocentric distance \(d\) far beyond the semi-major axes of the
well-measured planetary orbits, the leading perturbation it exerts on the inner Solar
System is tidal, governed by
\begin{equation}
  K_X
  \equiv
  \frac{G M_X}{d^3}.
  \label{eq:tidalK}
\end{equation}
An ephemeris constraint on this tidal field is therefore, in effect, an upper bound on
\(M_X/d^3\).

On the timescale of modern ephemeris data, such a companion would barely move. Its orbital
period is
\begin{equation}
  P_X
  \simeq
  \left(\frac{d}{1\,\AU}\right)^{3/2}{\rm yr}
  \simeq
  9\times10^4\,{\rm yr}
  \left(\frac{d}{2000\,\AU}\right)^{3/2},
\end{equation}
so over the few decades spanned by high-precision ranging data it acts, to excellent
approximation, as a static external tidal field. A full treatment would need to account for
sky direction, orbital phase, inclination, eccentricity, and correlations with the fitted
ephemeris parameters~\cite{Fienga:2020,HolmanPayne:2016}; we do not attempt that level of
detail here. Instead we adopt a single illustrative tidal envelope, calibrated against
existing Planet-Nine-like constraints, and use it as a stand-in for the genuine
direction-dependent bound.

We write
\begin{equation}
  M_{\rm max}^{\rm eph}(d)
  =
  Kd^3,
  \label{eq:Mmax_eph}
\end{equation}
with
\begin{equation}
  K\simeq 1.15\times10^{-13}
  \frac{\Msun}{\AU^3},
  \label{eq:Kvalue}
\end{equation}
calibrated to the INPOP19a constraints~\cite{Fienga:2020},
\begin{equation}
  M_X\simeq5\,\Mearth
  \quad \text{excluded at} \quad
  d\lesssim500\,\AU,
\end{equation}
and
\begin{equation}
  M_X\simeq10\,\Mearth
  \quad \text{excluded at} \quad
  d\lesssim650\,\AU.
\end{equation}
The two calibration points agree on \(M_X/d^3\) to within ten percent, which is sufficient
for the illustrative tidal envelope used here. Equation~\eqref{eq:Mmax_eph} should be
read as an illustrative, direction-averaged scaling, not as a substitute for a full
ephemeris likelihood. In Earth-mass units it reads
\begin{equation}
  M_{\rm max}^{\rm eph}(d)
  \simeq
  38\,\Mearth
  \left(
  \frac{d}{1000\,\AU}
  \right)^3.
  \label{eq:Mmax_eph_num}
\end{equation}

The \(d^3\) scaling in Eq.~\eqref{eq:Mmax_eph} can also be understood directly from the
leading tidal expansion. For a distant companion, the range perturbation induced by the
external field takes the schematic form
\begin{equation}
  \Delta \rho(t;\hat{\mathbf n},d,M_X)
  =
  K_X\,s(t;\hat{\mathbf n}),
  \qquad
  K_X=\frac{G M_X}{d^3},
  \label{eq:range_tidal_scaling}
\end{equation}
where \(s(t;\hat{\mathbf n})\) is the response to a unit tidal field in the direction
\(\hat{\mathbf n}\). Higher multipoles are suppressed by \(O(a_{\rm Sat}/d)\), about
\(2\%\) at \(d=500\,\AU\) and smaller beyond. Thus, at leading tidal order, any ephemeris
constraint on the external field has the form
\begin{equation}
  M_{\rm max}(d,\hat{\mathbf n})
  =
  \frac{K_{\rm max}(\hat{\mathbf n})d^3}{G}.
  \label{eq:Mmax_directional_tidal}
\end{equation}
The exponent \(d^3\) is therefore fixed by the quadrupolar tidal scaling; the nontrivial
information is the normalisation and the direction dependence of \(K_{\rm max}\).

As a check on the latter, we constructed a minimal projected range model in the ecliptic
plane, using circular Earth and Saturn orbits, the Earth--Saturn range as observable, and
projecting out the response to the in-plane initial conditions of both planets. In raw
form this simplified model is about \(30\text{--}40\) times more constraining than the
published INPOP-like Planet-Nine exclusions, reflecting the many additional nuisance
parameters, correlations, and systematics present in a real global ephemeris fit. We
therefore use it only for its direction-dependent shape, calibrating its absolute
normalisation to the two reference points \(5\,M_\oplus\) at \(500\,\AU\) and
\(10\,M_\oplus\) at \(650\,\AU\). The required calibration factors, \(36.2\) and \(32.9\),
are close, confirming that the simplified model has the correct \(d^3\) scaling. After
calibration, the ecliptic-longitude dependence changes the allowed mass by a factor of
order two around the median. At \(2000\,\AU\), for example, the calibrated range is
approximately
\begin{equation}
  M_{\rm max}(2000\,\AU)
  \simeq
  0.74\text{--}1.72\,\Mjup,
\end{equation}
with a median close to \(0.96\,\Mjup\). This supports the use of the scalar envelope
\eqref{eq:Mmax_eph} as an order-of-magnitude, direction-averaged guide, while making clear
that its absolute normalisation is imported from existing ephemeris studies rather than
derived from a full new fit.

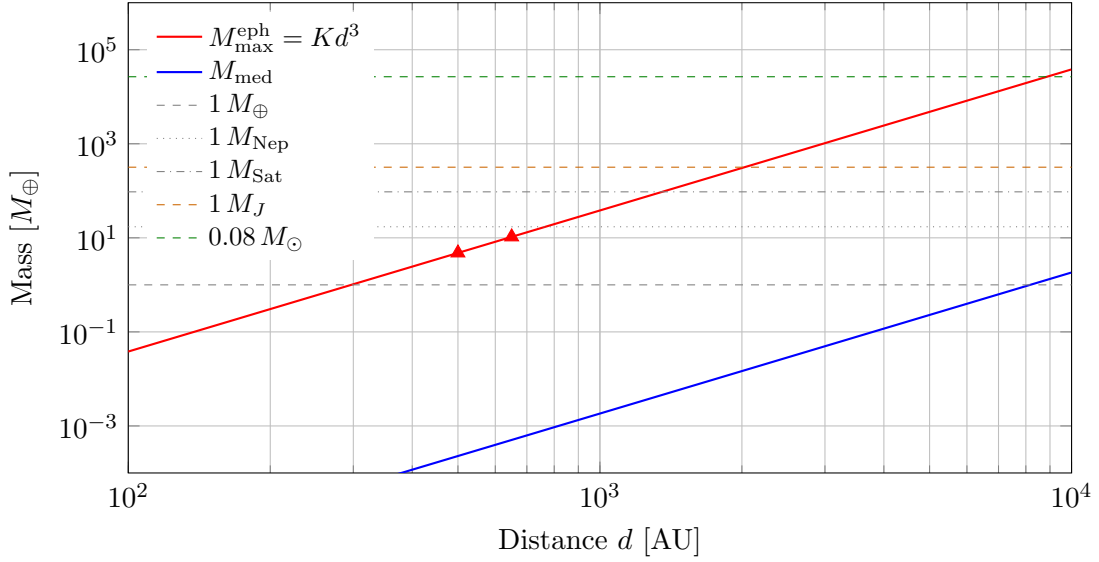
\begin{figure}[t]
\centering
\begin{tikzpicture}
\begin{loglogaxis}[
  width=0.90\textwidth,
  height=0.50\textwidth,
  xmin=100, xmax=10000,
  ymin=1e-4, ymax=1e6,
  xlabel={Distance \(d\) [\AU]},
  ylabel={Mass [\(\Mearth\)]},
  grid=both,
  legend style={at={(0.02,0.98)},anchor=north west,draw=none,fill=white,
                font=\small},
  legend cell align=left,
]
\addplot[red, thick, domain=100:10000, samples=200]
  {1.1473e-13*x^3/3.003e-6};
\addlegendentry{\(M_{\rm max}^{\rm eph}=Kd^3\)}
\addplot[blue, thick, domain=100:10000, samples=200]
  {6.044*0.008*(x/206265)^3/3.003e-6};
\addlegendentry{\(M_{\rm med}\)}
\addplot[gray, dashed,     domain=100:10000] {1};
\addlegendentry{\(1\,\Mearth\)}
\addplot[gray, dotted,     domain=100:10000] {17.2};
\addlegendentry{\(1\,\Mnep\)}
\addplot[gray, dashdotted, domain=100:10000] {95.1};
\addlegendentry{\(1\,\Msat\)}
\addplot[orange!80!black, dashed, domain=100:10000] {317.9};
\addlegendentry{\(1\,\Mjup\)}
\addplot[green!50!black,  dashed, domain=100:10000] {0.08/3.003e-6};
\addlegendentry{\(0.08\,\Msun\)}
\addplot[only marks, mark=triangle*, mark size=3pt, red]
  coordinates {(500,4.78) (650,10.49)};
\end{loglogaxis}
\end{tikzpicture}
\caption{%
Comparison between the smooth-halo median mass scale \(M_{\rm med}(d)\) (blue) and the
illustrative ephemeris tidal envelope (red). The blue curve gives the monochromatic object
mass for which the median nearest-neighbour distance in a smooth local halo is \(d\). The
red curve gives the illustrative ephemeris upper envelope
\(M_{\rm max}^{\rm eph}=Kd^3\), calibrated to Planet-Nine-like INPOP19a constraints
(filled triangles). The vertical gap shows that planetary or stellar hidden-brane objects
allowed by Solar-System dynamics are far too massive to be typical smooth-halo draws.
}
\label{fig:budget_ephemeris}
\end{figure}

Figure~\ref{fig:budget_ephemeris} sets this envelope against the smooth-halo median mass
scale from the previous section. The two curves are separated by several orders of
magnitude at every distance shown, and that gap is really the central result of this part
of the paper: whatever mass Solar-System dynamics still allows at a given distance, it is
far more than a smooth local halo could ever supply there.

Table~\ref{tab:crossings} lists the corresponding crossings of the ephemeris envelope.

\begin{table}[h]
\centering
\begin{tabular}{ccc}
\toprule
Distance \(d\) & \(M_{\rm max}^{\rm eph}(d)\) & Reference \\
\midrule
\(300\,\AU\)   & \(1.0\,\Mearth\)    & Earth mass \\
\(500\,\AU\)   & \(4.8\,\Mearth\)    & super-Earth \\
\(650\,\AU\)   & \(10.5\,\Mearth\)   & INPOP19a anchor \\
\(766\,\AU\)   & \(17\,\Mearth\)     & Neptune mass \\
\(1000\,\AU\)  & \(38\,\Mearth\)     & sub-Saturn \\
\(1355\,\AU\)  & \(95\,\Mearth\)     & Saturn mass \\
\(1500\,\AU\)  & \(129\,\Mearth\)    & sub-Jovian \\
\(2026\,\AU\)  & \(318\,\Mearth\)    & Jupiter mass \\
\bottomrule
\end{tabular}
\caption{Mass crossings of the illustrative ephemeris tidal envelope
\(M_{\rm max}^{\rm eph}=Kd^3\), \(K=1.15\times10^{-13}\,\Msun/\AU^3\). The table is meant
for the range where the envelope is used as a phenomenological guide; formal
extrapolations to several thousand AU should not be read as precise bounds.}
\label{tab:crossings}
\end{table}

If the same \(d^3\) envelope were extrapolated formally, its crossing with the ordinary
hydrogen-burning limit \(0.08\,\Msun\) would occur at
\(d\simeq8.9\times10^3\,\AU\). This number should not be read as an exclusion boundary.
The envelope is calibrated at \(500\text{--}650\,\AU\), and at several thousand AU
Galactic tides, stellar passages, and Oort-cloud dynamics become relevant
\cite{HeislerTremaine:1986,Duncan:1987,Dones:2004}. The extrapolated crossing is useful
only as an order-of-magnitude statement: ordinary stellar masses are pushed to Oort-cloud
scales.

Within this illustrative picture, a hidden-brane bound companion of super-Earth to Neptune
mass is allowed, at the level of this illustrative envelope, at \(300\text{--}800\,\AU\);
a sub-Saturn object at \(1000\,\AU\); Saturn to sub-Jovian masses by
\(1350\text{--}1500\,\AU\); and a Jovian one around \(2000\,\AU\). None of these is a
smooth-halo MACHO in the usual sense. Each is a structured hidden companion whose
constraints come from Solar-System dynamics, not from halo microlensing statistics.



\section{Population and dark-sector requirements}
\label{sec:population_dark_sector}

The previous sections establish a dynamical window for a hidden-brane companion, but they
say nothing about why such an object should exist in the first place. The smooth-halo
picture has already been set aside as a typical origin, which leaves the bound-companion
interpretation as the relevant one for the rest of this paper: a nearby hidden object must
come from a structured hidden sector, one capable of cooling, fragmenting, and assembling
bound objects rather than persisting as a featureless collisionless halo. We do not attempt
to build a complete model of hidden structure formation here; the aims of this section are
more modest. We first introduce a phenomenological population parameter that captures how
often hidden companions occur around ordinary stellar systems. We then work through a
simple QCD-scaled benchmark showing that the Jovian mass window already selected by
Solar-System dynamics can arise naturally once the hidden baryon mass is allowed to exceed
the ordinary proton mass.

\subsection{Population parameters}

The quantity of interest is not the fraction of the smooth Galactic halo locked up in
compact objects, but the probability that a given stellar system hosts a bound hidden
companion at all. A convenient way to express this is the differential parametrization
\begin{equation}
  \frac{d^2 f_{\rm host}}{d\ln M\,d\ln a}(M,a),
  \label{eq:fhost_differential}
\end{equation}
where \(M\) is the companion mass and \(a\) its semi-major axis. Thus
\begin{equation}
  df_{\rm host}
  =
  \frac{d^2 f_{\rm host}}{d\ln M\,d\ln a}
  d\ln M\,d\ln a
\end{equation}
is the fraction of stellar systems hosting a hidden companion in the corresponding
logarithmic interval of mass and semi-major axis. This is a genuinely different object
from the halo fraction \(f_B(M)\) of the previous section: \(f_B(M)\) measures how much of
the smooth Galactic halo is made of compact objects, while \(f_{\rm host}\) measures how
often a bound hidden companion turns up around a star.

We make no attempt to predict \(f_{\rm host}\) here. What the present analysis offers
instead is the region of the \((M,a)\) plane where such companions are not already excluded
by Solar-System dynamics. A point of language is worth flagging: the ephemeris constraints
of the previous section are phrased in terms of the instantaneous heliocentric distance
\(d\), whereas a population model is more naturally phrased in terms of orbital elements
such as \(a\), eccentricity, inclination, and orbital phase. In what follows we identify
\(d\) with the companion's characteristic orbital scale, with the understanding that a full
treatment would require the orbit-dependent ephemeris likelihood rather than this
shorthand.

In a dissipative hidden sector, \(f_{\rm host}\) would depend on a chain of model-dependent
ingredients: the fraction of dark matter in the dissipative component, the hidden cooling
rate, the hidden angular momentum distribution, the efficiency of fragmentation, and
whatever history of capture or co-formation linked the hidden object to its visible
host~\cite{Fan:2013yva,Buckley:2017ijx,Cirelli:2024ssm}. These ingredients can give rise
to several qualitatively different kinds of hidden companion:
\begin{enumerate}
  \item A \emph{dark compact remnant}, supported by degeneracy pressure or by some other
        hidden-sector equation of state, with no ongoing hidden nuclear burning.
  \item A \emph{dark planet}, formed by cooling and fragmentation of hidden matter below
        the threshold needed for sustained burning.
  \item A \emph{mirror star} or \emph{mirror planet}, in a hidden sector whose atomic and
        nuclear microphysics mirrors that of the Standard Model.
  \item A \emph{dark star with active burning}, requiring hidden nuclear reactions or some
        other long-lived internal energy source.
\end{enumerate}
The remainder of this section focuses on the last of these, not because it is the most
likely outcome, but because it offers the simplest benchmark for the characteristic mass
scale involved.



\subsection{A QCD-scaled hidden stellar benchmark}

The question worth asking is whether the mass window already selected by the ephemeris
bound can overlap with the minimum mass for a genuine hidden star. In a simple QCD-scaled
benchmark, it can.

Consider a hidden sector with Standard-Model-like atomic and nuclear structure, but with a
larger confinement scale,
\begin{equation}
  \Lambda'_{\rm QCD}
  =
  \kappa\,\Lambda_{\rm QCD},
  \qquad
  \kappa>1,
  \label{eq:qcd_scaled_kappa}
\end{equation}
so that the hidden baryon mass scales approximately as
\begin{equation}
  m_{B'}\simeq \kappa m_p.
  \label{eq:hidden_baryon_mass}
\end{equation}
As a benchmark we take
\begin{equation}
  m_{e'}\sim m_e,
  \qquad
  \alpha'\sim\alpha.
  \label{eq:dark_alpha_benchmark}
\end{equation}
The second of these choices matters more than it might first appear. The hidden
electromagnetic coupling \(\alpha'\) governs hidden atomic physics, radiative opacity, and
cooling, but it has nothing to do with whether the object is visible to us. A hidden star
can be bright in hidden photons and still be completely dark on our brane, provided the
kinetic mixing between the hidden and visible photons is small~\cite{Holdom:1985ag},
\begin{equation}
  \varepsilon_{\gamma\gamma'}\ll1.
  \label{eq:kinetic_mixing_small}
\end{equation}
Keeping \(\alpha'\sim\alpha\) therefore preserves a recognisable hidden atomic and radiative
physics, while \(\varepsilon_{\gamma\gamma'}\simeq0\) is what actually keeps the object
invisible to ordinary telescopes.

The same language naturally addresses the stability and relic abundance of the hidden
baryons. Suppose the hidden sector carries an exactly conserved, or at least sufficiently
long-lived, baryon number \(B'\); the lightest hidden baryon is then cosmologically stable.
After hidden confinement, the symmetric \(B'\bar B'\) component can annihilate away
efficiently into lighter hidden states, leaving behind an abundance set by a hidden baryon
asymmetry rather than by a conventional thermal freeze-out, as in asymmetric-dark-matter
scenarios~\cite{Kaplan:2009ag,Petraki:2013wwa}. Writing \(T'=\xi T\) for the hidden-sector
temperature and defining
\begin{equation}
  \eta_{B'}
  \equiv
  \frac{n_{B'}-n_{\bar B'}}{n_{\gamma'}},
\end{equation}
the hidden baryon abundance relative to the visible one is, parametrically,
\begin{equation}
  \frac{\Omega_{B'}}{\Omega_b}
  \simeq
  \frac{m_{B'}}{m_p}
  \frac{\eta_{B'}}{\eta_b}
  \xi^3
  \simeq
  \kappa\,
  \frac{\eta_{B'}}{\eta_b}
  \xi^3.
  \label{eq:hidden_asymmetry_abundance}
\end{equation}
If the dissipative hidden sector accounts for only a fraction \(f_{\rm diss}\) of the total
dark matter, \(\Omega_{B'}=f_{\rm diss}\,\Omega_{\rm DM}\), then, using the observed ratio
\(\Omega_{\rm DM}/\Omega_b\simeq5.3\), this becomes
\begin{equation}
  \kappa\,
  \frac{\eta_{B'}}{\eta_b}
  \xi^3
  \simeq
  f_{\rm diss}
  \frac{\Omega_{\rm DM}}{\Omega_b}
  \simeq
  5.3\,f_{\rm diss}.
  \label{eq:fdiss_relation}
\end{equation}
For \(\kappa\sim10\), an order-one hidden baryon asymmetry relative to the visible one is
therefore enough to account for an order-one dark matter fraction, provided the hidden
sector sits at a comparable temperature. If the dissipative component is instead only a
subdominant piece of the dark matter, the asymmetry required shrinks accordingly. This is a
useful point to have on record: forming a bound hidden companion calls for a dissipative
component large enough to cool and fragment, but it does not require that the entire dark
matter budget reside there.

That the dissipative component can be subdominant matters for more than just the
asymmetry bookkeeping above. Scenarios in which a large fraction of the dark matter is
strongly self-interacting or dissipative are subject to a range of cosmological and
astrophysical constraints, from halo shapes, cluster collisions, dark acoustic oscillations
and dark-disk formation to the broader pattern of small-scale structure
\cite{Cirelli:2024ssm,Buckley:2017ijx,CyrRacine:2015ihg}. These constraints are,
however, constraints on a cosmological population: they depend on the abundance,
temperature, couplings, and cooling history of the interacting component taken as a whole.
The scenario considered here asks for less. It requires a hidden component capable of
cooling and occasionally forming a bound companion, not a fully dissipative Galactic halo.
We therefore treat these global bounds as restrictions on which hidden-sector embeddings
remain viable, rather than as direct constraints on the local mass-distance window studied
here.

A colder hidden sector, \(\xi<1\), or a later entropy transfer into the visible sector,
could also help satisfy dark-radiation constraints if the hidden photon stays light; we do
not model that cosmological history here. The estimate above is meant only to establish
that a QCD-scaled hidden baryon sector can carry a stable relic component of the right
order of magnitude, without appealing to a WIMP-like freeze-out.

This benchmark falls well short of a complete hidden stellar model. A genuine calculation
would need the hidden nuclear binding energies, a hidden analogue of the proton-proton
chain or some other energy-generating reaction, the hidden opacity, the hidden photon
temperature, and the cosmological abundance of the dissipative component. We use it only
to fix a parametric mass scale.

The relevant scaling is the familiar one: characteristic stellar and degenerate-object
masses go as
\begin{equation}
  M_\star \sim \frac{M_{\rm Pl}^3}{m_B^2}.
\end{equation}
With \(m_{B'}\simeq\kappa m_p\), and normalising to the ordinary
hydrogen-burning threshold~\cite{Burrows:2001}, the minimum mass for sustained hidden
burning scales as
\begin{equation}
  M_{\rm min}^{\rm burn'}(\kappa)
  \simeq
  \frac{0.08\,\Msun}{\kappa^2},
  \label{eq:Mmin_kappa}
\end{equation}
and the same scaling lowers the Chandrasekhar mass of hidden degenerate remnants,
\begin{equation}
  M_{\rm Ch}'(\kappa)
  \simeq
  \frac{1.4\,\Msun}{\kappa^2}.
  \label{eq:MCh_kappa}
\end{equation}
Both expressions should be read as scaling relations rather than precision predictions:
dimensionless nuclear, opacity, and equation-of-state factors can shift the numerical
coefficients by some amount.

For \(\kappa=10\), the benchmark gives
\begin{equation}
  M_{\rm min}^{\rm burn'}
  \simeq
  8\times10^{-4}\,\Msun
  \simeq
  0.8\,\Mjup,
  \qquad
  M_{\rm Ch}'
  \simeq
  1.4\times10^{-2}\,\Msun
  \simeq
  15\,\Mjup.
  \label{eq:kappa10_masses}
\end{equation}
A moderately enhanced hidden confinement scale moves the characteristic hidden stellar mass
down from the ordinary hydrogen-burning scale into the Jovian range. The hidden
Chandrasekhar mass for the same benchmark sits considerably higher, around \(15\,\Mjup\),
so the ephemeris window near \(d\sim2000\,\AU\)---which permits roughly one Jupiter
mass---selects objects well below the hidden Chandrasekhar scale. A hidden companion near
the burning threshold would therefore be a very low-mass active hidden star rather than a
Chandrasekhar-scale remnant. This does not rule out colder or partially degenerate objects
of lower mass; it simply shows that the active-burning threshold can land naturally inside
the ephemeris-allowed window.

Combining Eq.~\eqref{eq:Mmin_kappa} with the illustrative ephemeris envelope
\(M_{\rm max}^{\rm eph}(d)=Kd^3\) gives the value of \(\kappa\) needed for the hidden
burning threshold to fall below the ephemeris bound,
\begin{equation}
  \kappa_{\rm req}(d)
  =
  \left[
  \frac{0.08\,\Msun}{Kd^3}
  \right]^{1/2}
  \propto d^{-3/2}.
  \label{eq:kappa_req}
\end{equation}
Table~\ref{tab:kappa_req} gives the numbers.

\begin{table}[h]
\centering
\begin{tabular}{cccc}
\toprule
Distance \(d\) & \(M_{\rm max}^{\rm eph}\,[M_\oplus]\) &
\(M_{\rm max}^{\rm eph}\,[\Mjup]\) & \(\kappa_{\rm req}\) \\
\midrule
\(1000\,\AU\) &  38 & 0.12 & 26 \\
\(1355\,\AU\) &  95 & 0.30 & 17 \\
\(1500\,\AU\) & 129 & 0.41 & 14 \\
\(2000\,\AU\) & 306 & 0.96 &  9 \\
\(2500\,\AU\) & 597 & 1.88 &  7 \\
\bottomrule
\end{tabular}
\caption{Required confinement enhancement \(\kappa_{\rm req}(d)\) for the hidden burning
threshold to lie below the illustrative ephemeris envelope, at selected distances.}
\label{tab:kappa_req}
\end{table}

The most interesting region is therefore
\begin{equation}
  d\sim1500\text{--}2000\,\AU,
  \qquad
  \kappa\sim10\text{--}15,
  \label{eq:interesting_kappa_window}
\end{equation}
where the ephemeris envelope allows sub-Jovian to Jovian masses and the QCD-scaled burning
threshold can sit in the same range. In this window, the hidden companion is not
necessarily a cold object: it can be a genuine hidden-sector star, bright in hidden photons
yet visible to us only through gravity.

\begin{figure}[t]
\centering
\begin{tikzpicture}
\begin{semilogxaxis}[
  width=0.95\textwidth,
  height=0.40\textwidth,
  xmin=800, xmax=3000,
  ymin=5, ymax=35,
  xlabel={Distance \(d\) [\AU]},
  ylabel={Required \(\kappa=\Lambda'_{\rm QCD}/\Lambda_{\rm QCD}\)},
  grid=both,
  legend style={at={(0.98,0.98)},anchor=north east,draw=none,fill=white,
                font=\small},
  legend cell align=left,
]
\addplot[red, thick, domain=800:3000, samples=200]
  {sqrt(0.08/(1.1473e-13*x^3))};
\addlegendentry{\(\kappa_{\rm req}(d)\)}
\addplot[gray, dashed, domain=800:3000] {10};
\addlegendentry{\(\kappa=10\)}
\addplot[gray, dotted, domain=800:3000] {15};
\addlegendentry{\(\kappa=15\)}
\addplot[only marks, mark=*, mark size=2.5pt, red]
  coordinates {(1000,26.4) (1500,14.4) (2000,9.3) (2500,6.7)};
\end{semilogxaxis}
\end{tikzpicture}
\caption{%
Value of \(\kappa=\Lambda'_{\rm QCD}/\Lambda_{\rm QCD}\) required for the hidden-sector
burning threshold \(M_{\rm min}^{\rm burn'}\simeq0.08\,\Msun/\kappa^2\) to lie below the
illustrative ephemeris mass envelope \(M_{\rm max}^{\rm eph}(d)=Kd^3\). The range
\(d\sim1500\text{--}2000\,\AU\) corresponds to \(\kappa\sim10\text{--}15\), a moderate
enhancement of the hidden confinement scale.
}
\label{fig:kappa_req}
\end{figure}
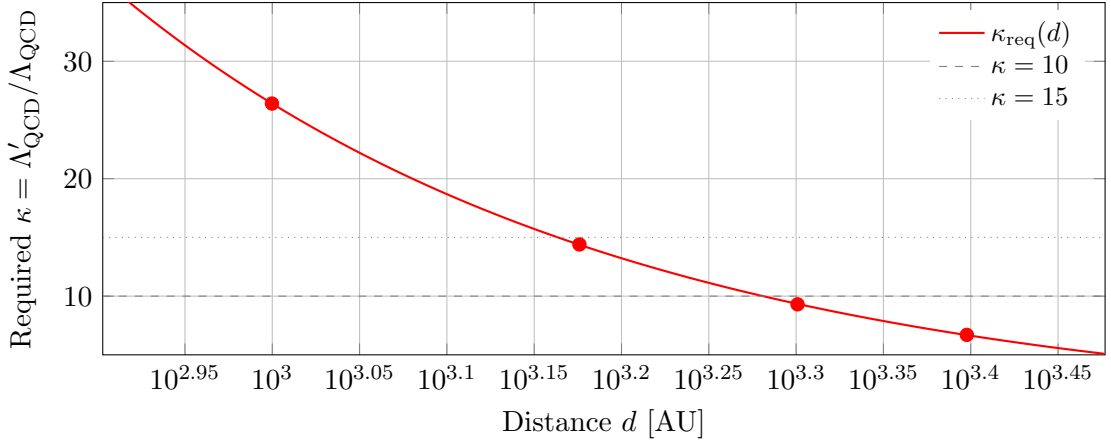

Below the hidden hydrogen-burning threshold, the same sector can still form dark planets,
cold partially degenerate objects, or hidden brown dwarfs. If the hidden deuterium-burning 
threshold follows the same scaling, one expects
parametrically, again normalising to the ordinary substellar threshold~\cite{Burrows:2001},
\begin{equation}
  M_D'(\kappa)
  \sim
  \frac{13\,\Mjup}{\kappa^2}.
\end{equation}
For \(\kappa=10\), this places the hidden brown-dwarf interval roughly between
\(M_D'\sim0.13\,\Mjup\) and \(M_{\rm min}^{\rm burn'}\sim0.8\,\Mjup\), modulo the
uncertainties already inherent in the hidden nuclear physics. Objects lighter than this
range would behave more like dark planets or cold substellar bodies than like stars in any
sense.

The assumption \(\alpha'\sim\alpha\) is what makes hidden atomic cooling, molecular
physics, and radiative opacity plausible in the first place, while
\(\varepsilon_{\gamma\gamma'}\simeq0\) is what keeps the entire system electromagnetically
dark on our brane. The QCD-scaled benchmark should be read in that spirit: as an existence
proof for the relevant mass scale, not as a unique or preferred model of hidden stellar
structure.



\section{Source-strength proxy and the need for time dependence}
\label{sec:source_strength_proxy}

The previous section showed that a hidden companion in the ephemeris-allowed window need
not be a cold, inert body: it can in principle be a genuine hidden-sector star. That
matters, but it is not by itself enough for a brane-to-brane gravitational signal of the
kind considered in the companion paper~\cite{Benakli:2026qag}, because mass alone does not
produce one. A given mass distribution sources the ordinary four-dimensional Newtonian
field through the massless graviton regardless of what it is doing, but exciting the
propagating Kaluza--Klein tower asks for something more specific: stress-energy that
varies in time, and varies fast enough.

The existence window derived above does not depend on the signal estimates in this
section. The purpose here is narrower: to identify the parametric requirements that would
have to be met if the hidden companion were used as a source or target for the
brane-to-brane channel discussed in Ref.~\cite{Benakli:2026qag}. This section works out
where that threshold sits, why ordinary Solar-System motion falls hopelessly short of it,
and what is left over once the easy possibilities are excluded. It ends with a
source-strength proxy, defined carefully enough not to be mistaken for more than it is.

\subsection{What controls the KK response}

We assume, for definiteness, a single compact extra dimension of size \(L\), giving a
discrete Kaluza--Klein spectrum stacked above the massless four-dimensional graviton. With
the same boundary-condition convention as in Ref.~\cite{Benakli:2026qag}, the first
threshold is
\begin{equation}
  k_1\simeq \frac{\pi}{L},
  \qquad
  \omega_1\simeq \frac{\pi c}{L},
  \qquad
  f_1=\frac{\omega_1}{2\pi}\simeq \frac{c}{2L}.
  \label{eq:first_KK_threshold}
\end{equation}
For \(L=0.1\,\mu{\rm m}\), this gives
\begin{equation}
  m_1 c^2
  \simeq
  6\,{\rm eV},
  \qquad
  \omega_1
  \simeq
  9.4\times10^{15}\,{\rm rad/s},
  \qquad
  f_1
  \simeq
  1.5\times10^{15}\,{\rm Hz}.
  \label{eq:KK_scales_01um}
\end{equation}
This treats the first KK threshold as discrete; with several large extra dimensions the
spectrum would be denser, and the relevant quantity would be a density of accessible modes
rather than a single isolated threshold. It is worth being upfront about where this leaves
us: we do not assume any detector capable of measuring gravitational KK radiation near
\(10^{15}\,{\rm Hz}\). Everything that follows in this section is phenomenological in a
stronger sense than the rest of the paper---it locates where a signal would have to live,
not whether anyone could measure it. We come back to this point at the end.

Two effects are easy to run together, and it is worth keeping them apart. A static source
simply does not radiate: even the massless four-dimensional graviton needs time-dependent
stress-energy to produce a propagating wave, and a stationary configuration gives only a
static potential. Separately, and on top of that, a massive KK mode imposes its own
threshold. For a source component oscillating at angular frequency \(\omega\), the wave
number of the \(n\)-th KK mode is
\begin{equation}
  k_n(\omega)
  =
  \frac{1}{c}
  \sqrt{\omega^2-\omega_n^2},
  \qquad
  \omega_n=\frac{m_n c^2}{\hbar}.
  \label{eq:kn_dimensional}
\end{equation}
Above threshold, \(\omega>\omega_n\), the mode propagates with the usual oscillatory
\(1/d\) falloff, supplemented near threshold by \(k_n\)-dependent factors. Below threshold,
\(\omega<\omega_n\), the response is evanescent and dies exponentially,
\begin{equation}
  G_n(d;\,\omega<\omega_n)
  \sim
  \frac{
  \exp\!\left[-\sqrt{\omega_n^2-\omega^2}\,d/c\right]
  }{d}.
  \label{eq:yukawa}
\end{equation}
So the requirements stack: the source first needs genuine time dependence, and that time
dependence then needs to clear the KK threshold before the mode will propagate at all.


\subsection{Why macroscopic motion cannot excite the KK tower}

Orbital motion is nowhere close. At \(d\simeq2000\,\AU\),
\begin{equation}
  f_{\rm orb}
  \simeq
  3.5\times10^{-13}\,{\rm Hz}
  \left(\frac{d}{2000\,\AU}\right)^{-3/2},
  \label{eq:orbital_frequency_2000}
\end{equation}
which sits twenty-eight orders of magnitude below \(f_1\simeq1.5\times10^{15}\,{\rm Hz}\).

Letting the companion's own internal structure oscillate does much better, though still
not nearly enough. The dynamical frequency of a self-gravitating body scales with the
square root of its mean density,
\begin{equation}
  f_{\rm dyn}
  \sim
  \frac{1}{2\pi}\sqrt{G\bar{\rho}}.
\end{equation}
Jupiter (\(\bar{\rho}\sim10^3\,{\rm kg/m^3}\)) gives \(f_{\rm dyn}\sim10^{-4}\,{\rm Hz}\); a
white dwarf (\(\bar{\rho}\sim10^9\,{\rm kg/m^3}\)) gives
\(f_{\rm dyn}\sim4\times10^{-2}\,{\rm Hz}\); a neutron star
(\(\bar{\rho}\sim10^{17}\,{\rm kg/m^3}\)) gives \(f_{\rm dyn}\sim4\times10^2\,{\rm Hz}\).
Even pushing the QCD-scaled hidden remnant of Section~\ref{sec:population_dark_sector} to
\(\kappa^4\) times nuclear density for \(\kappa=10\) only reaches
\(f_{\rm dyn}\sim4\times10^4\,{\rm Hz}\), still about eleven orders of magnitude short of
\(f_1\). Closing that remaining gap through density alone would need
\(\bar{\rho}\sim10^{42}\,{\rm kg/m^3}\), about twenty-five orders of magnitude above
nuclear density---not a configuration any macroscopic self-gravitating object can reach.

The conclusion holds without qualification: no macroscopic motion of a companion at
Solar-System scale, however compact, gets anywhere near a KK threshold of
\(f_1\simeq1.5\times10^{15}\,{\rm Hz}\). The companion supplies a large mass-energy at a
favourable distance, but its motion alone does not make it a KK transmitter.



\subsection{Microscopic excitation and what \(\epsilon_n(\omega)\) parametrises}

If the KK tower is to be excited at all, the stress-energy doing the exciting has to live
at the microscopic level. For \(L=0.1\,\mu{\rm m}\), \(\omega\gtrsim m_1c^2/\hbar\)
corresponds to a threshold energy \(m_1c^2\simeq6\,{\rm eV}\), i.e.\ ultraviolet
frequencies for the first KK mode, and to progressively higher ultraviolet, X-ray, or
gamma-ray frequencies for the modes above it. Hidden atomic, plasma, molecular, or nuclear
processes naturally involve such frequencies. This statement, however, is purely
kinematic. Atomic or nuclear transitions primarily emit hidden photons or other
hidden-sector quanta; they do not by themselves constitute an efficient graviton or
KK-graviton source. What matters for the KK channel is the part of the hidden
stress-energy tensor that varies coherently at the relevant frequency and has the
appropriate multipolar structure to couple to the gravitational mode.

This is precisely the hard part. Random microscopic events do not generally add
constructively as a gravitational source: even if the hidden star contains many processes
with energies above the KK threshold, their phases, locations, and multipole moments may
average out, leaving only a tiny coherent component of the total stress-energy. A plasma
mode, collective excitation, or phase transition could in principle produce a larger
coherent component, but estimating that would require a concrete model of the hidden
stellar interior, which we do not attempt here.

We make no attempt to estimate this microphysics here, and instead fold the uncertainty
into a single dimensionless coefficient \(\epsilon_n(\omega)\). A schematic received
signal scale for the \(n\)-th mode can then be written as
\begin{equation}
  \mathcal{S}_n(M,r;\omega)
  \sim
  \epsilon_n(\omega)\,
  \mathcal{T}_n(\omega)\,
  \frac{GM}{c^2}\,
  G_n(\omega,r).
  \label{eq:Sn_general}
\end{equation}
Here \(r\) is the ordinary three-dimensional separation between the hidden source and the
receiver on our brane. The factor \(\mathcal{T}_n(\omega)\) denotes the compact-direction
transfer factor: the mode wave-function overlap with the source and detector brane
positions, together with any ultraviolet or brane-structure form factors. It does not
include the ordinary three-dimensional propagation factor, which is written separately as
\(G_n(\omega,r)\).

In the propagating regime, sufficiently far above threshold and away from near-field
effects, the four-dimensional KK Green function has the usual spherical-wave scaling,
\begin{equation}
  G_n(\omega,r)
  \sim
  \frac{e^{ik_n r}}{r},
  \qquad
  \omega>\omega_n.
  \label{eq:Gn_propagating}
\end{equation}
Below threshold, the same factor is evanescent,
\begin{equation}
  G_n(\omega,r)
  \sim
  \frac{
  \exp\!\left[-\sqrt{\omega_n^2-\omega^2}\,r/c\right]
  }{r},
  \qquad
  \omega<\omega_n.
  \label{eq:Gn_evanescent}
\end{equation}
Thus the \(1/r\) factor used below is not a five-dimensional Newtonian potential. It is
the ordinary three-dimensional propagation dilution of a four-dimensional KK mode after
dimensional reduction. In the Earth-based estimates of this section, \(r\simeq d\), since
the companion is at heliocentric distance \(d\gg1\,\AU\). Section~\ref{sec:local_transmission}
then asks what changes when the receiver is moved from Earth to a probe at separation
\(r=b\ll d\).

About all that can be said model-independently about the excitation efficiency is the bound
\begin{equation}
  0\le\epsilon_n(\omega)\le 1,
\end{equation}
since no excitation mechanism can place more coherent time-dependent stress-energy at the
relevant frequency than the source's total rest mass-energy. The limiting value
\(\epsilon_n=1\) would correspond to the entire source oscillating coherently as one; real
values are presumably far smaller, and depend on microphysics this paper does not attempt
to model.



\subsection{A source-scale proxy, honestly labelled}

The previous subsection separated three ingredients in the schematic signal scale:
the source strength \(GM/c^2\), the compact-direction transfer factor
\(\mathcal{T}_n(\omega)\), and the ordinary three-dimensional propagation factor
\(G_n(\omega,r)\). In the propagating regime, sufficiently far above threshold and away
from near-field effects,
\begin{equation}
  G_n(\omega,r)\sim \frac{e^{ik_n r}}{r},
\end{equation}
so an Earth-based received source scale is proportional to
\begin{equation}
  \frac{GM}{c^2}\,G_n(\omega,d)
  \sim
  \frac{GM}{c^2 d},
  \label{eq:earth_based_proxy}
\end{equation}
where \(d\) is the heliocentric distance of the companion and we have used
\(d\gg1\,\AU\) to identify it with the source--Earth separation. This is the proxy used
through the rest of this section. It is dimensionless and model-independent only at the
limited level intended here: it is the gravitational source scale multiplied by the
ordinary three-dimensional propagation dilution to an Earth-based receiver. It is not, by
itself, a predicted detector amplitude, and certainly not a flux. Any real signal would
still have to include the microphysical efficiency \(\epsilon_n(\omega)\), the
compact-direction transfer factor \(\mathcal{T}_n(\omega)\), the phase and threshold
structure of \(G_n(\omega,r)\), and the detector response.

With that scale in hand, we can ask which ephemeris-allowed companion gives the largest
Earth-based proxy. For an object saturating the illustrative ephemeris envelope,
\begin{equation}
  M=M_{\rm max}^{\rm eph}(d)=Kd^3,
\end{equation}
one finds
\begin{equation}
  \mathcal{S}_{\rm max}^{\oplus}(d)
  \sim
  \frac{G M_{\rm max}^{\rm eph}(d)}{c^2 d}
  =
  \frac{G K d^2}{c^2}.
  \label{eq:Smax_d2}
\end{equation}
The Earth-based proxy grows with distance rather than shrinking because the largest
ephemeris-allowed mass grows as \(d^3\), while ordinary propagation contributes only one
power of \(1/d\) to the received amplitude scale. Thus, within the range where the tidal
envelope is a reasonable guide, the most favourable source scale comes not from the nearest
allowed companion but from the most massive one still permitted at the largest dynamically
relevant distance. Normalising to the Neptune-mass crossing,
\begin{equation}
  \widehat{\mathcal{S}}_{\oplus}(d)
  =
  \frac{M_{\rm max}^{\rm eph}(d)/d}
       {\Mnep/766\,\AU}
  =
  \left(
  \frac{d}{766\,\AU}
  \right)^2,
  \label{eq:Shat}
\end{equation}
gives
\begin{equation}
  \widehat{\mathcal{S}}_{\oplus}(1500\,\AU)
  \simeq 3.8,
  \qquad
  \widehat{\mathcal{S}}_{\oplus}(2000\,\AU)
  \simeq 6.8.
  \label{eq:Shat_values}
\end{equation}
A companion sitting near the Jovian crossing is therefore several times more favourable,
in this Earth-based proxy, than a Neptune-mass one at \(766\,\AU\).

It is worth closing with the caveat that has been sitting underneath this whole discussion.
Even a large gain in \(\widehat{\mathcal{S}}_{\oplus}\) does nothing on its own unless
three further conditions are met: a microscopic excitation mechanism must populate the
relevant KK frequency with coherent stress-energy of non-negligible efficiency
\(\epsilon_n(\omega)\); the compact-direction transfer factor
\(\mathcal{T}_n(\omega)\) must be non-negligible; and a detector capable of responding to
the relevant KK frequency range would be required. We do not assume such a detector here.
The proxy should therefore be read not as a claim about detectability, but as a diagnostic
of which part of the ephemeris-allowed window would provide the largest Earth-based
received source scale if the remaining factors were ever to cooperate.

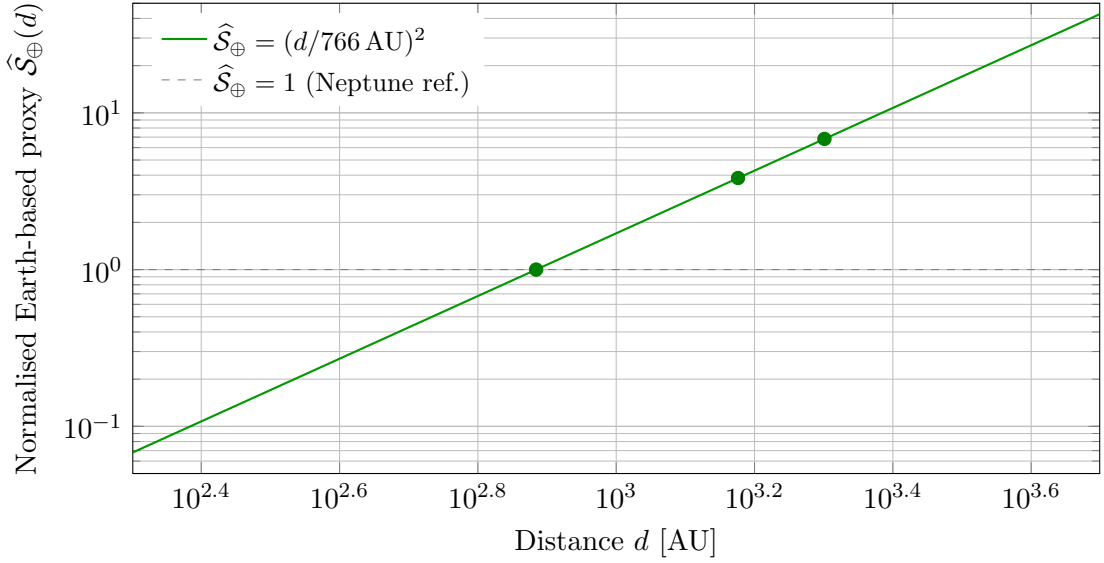
\begin{figure}[t]
\centering
\begin{tikzpicture}
\begin{loglogaxis}[
  width=0.92\textwidth,
  height=0.50\textwidth,
  xmin=200, xmax=5000,
  ymin=0.05, ymax=50,
  xlabel={Distance \(d\) [\AU]},
  ylabel={Normalised Earth-based proxy \(\widehat{\mathcal{S}}_{\oplus}(d)\)},
  grid=both,
  legend style={at={(0.02,0.98)},anchor=north west,draw=none,fill=white,
                font=\small},
  legend cell align=left,
]
\addplot[green!60!black, thick, domain=200:5000, samples=200]
  {(x/766)^2};
\addlegendentry{\(\widehat{\mathcal{S}}_{\oplus}=(d/766\,\AU)^2\)}
\addplot[gray, dashed, domain=200:5000] {1};
\addlegendentry{\(\widehat{\mathcal{S}}_{\oplus}=1\) (Neptune ref.)}
\addplot[only marks, mark=*, mark size=2.5pt, green!50!black]
  coordinates {(766,1.0) (1500,3.84) (2000,6.82)};
\end{loglogaxis}
\end{tikzpicture}
\caption{%
Normalised Earth-based source proxy
\(\widehat{\mathcal{S}}_{\oplus}(d)=(d/766\,\AU)^2\) for an object saturating the
illustrative ephemeris tidal envelope. Since
\(M_{\rm max}^{\rm eph}\propto d^3\), the source strength grows as \(d^3\), while the
ordinary three-dimensional propagation factor to an Earth-based receiver contributes one
power of \(1/d\) in the propagating regime. The result is
\(\widehat{\mathcal{S}}_{\oplus}\propto d^2\). The reference is a Neptune-mass hidden
companion at \(766\,\AU\). A companion near \(2000\,\AU\) is approximately seven times more
favourable in this proxy. This reflects available Earth-based received source scale only,
not detectability: the actual signal requires a microscopic excitation mechanism at the
relevant KK frequency, a non-negligible compact-direction brane-to-brane transfer factor,
and a detector response at that frequency.
}
\label{fig:signal_proxy}
\end{figure}



\section{Local transmission from the vicinity of the hidden companion}
\label{sec:local_transmission}

The previous section left two problems standing: the frequency a hidden source would need
to reach, and the coherence it would need to sustain there. The strategy discussed here
solves neither. What it addresses is a third, more mundane problem: the ordinary
three-dimensional propagation dilution between the hidden companion and an Earth-based
receiver.

Suppose the companion sits at heliocentric distance \(D\sim10^3\text{--}10^4\,\AU\). In
the notation of Eq.~\eqref{eq:Sn_general}, an Earth-based receiver samples the KK Green
function at an ordinary source--receiver separation \(r\simeq D\). For a propagating KK
mode, sufficiently far above threshold,
\begin{equation}
  G_n(\omega,r)\sim \frac{e^{ik_n r}}{r},
\end{equation}
so the Earth-based amplitude scale carries the geometric factor
\begin{equation}
  \mathcal{A}_\oplus \propto |G_n(\omega,D)|\sim \frac{1}{D}.
\end{equation}
At \(D\simeq2000\,\AU\) this penalty is severe. But once the companion's position has been
pinned down dynamically, there is a natural alternative: send a probe toward its
three-dimensional gravitational projection on our brane, and operate a receiver,
transmitter, or relay from close by, at ordinary spatial separation \(b\ll D\).

It is worth being precise about what such a move actually shortens. The probe does not
change the separation between the branes in the compact direction. That separation is a
property of the higher-dimensional geometry and enters the compact-direction transfer
factor \(\mathcal{T}_n(\omega)\): the KK wave-function overlap with the source and detector
brane positions, together with any ultraviolet or brane-structure form factors. Since the
probe remains on our brane, it samples the same compact-direction overlap as an
Earth-based receiver. What changes is only the ordinary three-dimensional separation
parallel to the branes, namely the argument \(r\) of \(G_n(\omega,r)\). The Earth-based
configuration has \(r\simeq D\), while the probe-based configuration has \(r=b\).

With that understood, the amplitude gain over Earth-based operation is
\begin{equation}
  \frac{\mathcal{A}_{\rm probe}}{\mathcal{A}_\oplus}
  \sim
  \frac{|G_n(\omega,b)|}{|G_n(\omega,D)|}.
  \label{eq:local_gain_general}
\end{equation}
For a propagating mode in the far-field regime, \(G_n(\omega,r)\sim e^{ik_n r}/r\), this
becomes
\begin{equation}
  \frac{\mathcal{A}_{\rm probe}}{\mathcal{A}_\oplus}
  \sim
  \frac{D}{b},
  \label{eq:local_gain_amp}
\end{equation}
and the corresponding flux or power gain follows as its square,
\begin{equation}
  \frac{\mathcal{F}_{\rm probe}}{\mathcal{F}_\oplus}
  \sim
  \left(
  \frac{\mathcal{A}_{\rm probe}}{\mathcal{A}_\oplus}
  \right)^2
  \sim
  \left(\frac{D}{b}\right)^2,
  \label{eq:local_gain_flux}
\end{equation}
assuming source and detector response are otherwise unchanged. For a mode below threshold,
the same statement is even stronger in principle, because the Green function is evanescent,
\begin{equation}
  G_n(\omega,r)
  \sim
  \frac{
  \exp\!\left[-\sqrt{\omega_n^2-\omega^2}\,r/c\right]
  }{r},
  \qquad
  \omega<\omega_n,
\end{equation}
so reducing \(r\) can remove not only a power-law dilution but also part of an exponential
suppression. We do not pursue that case explicitly; the point here is only that moving the
receiver close to the companion improves the ordinary propagation factor whenever a
brane-to-brane channel is present.

For a companion at \(D=2000\,\AU\), the propagating-mode gains are
\begin{align}
  b&=10\,\AU
  &&\Rightarrow\quad
  \mathcal{A}_{\rm gain}\sim200,
  &&\mathcal{F}_{\rm gain}\sim4\times10^4,
  \\
  b&=1\,\AU
  &&\Rightarrow\quad
  \mathcal{A}_{\rm gain}\sim2000,
  &&\mathcal{F}_{\rm gain}\sim4\times10^6,
  \\
  b&=0.1\,\AU
  &&\Rightarrow\quad
  \mathcal{A}_{\rm gain}\sim2\times10^4,
  &&\mathcal{F}_{\rm gain}\sim4\times10^8.
\end{align}
None of this implies the signal is detectable. It only shows that, if a propagating
brane-to-brane channel exists, operating near the companion's gravitational projection
beats operating from Earth by these factors, purely through the ordinary
three-dimensional Green function.

Whether this is feasible comes down to travel time. Witten has already proposed sending
small, fast spacecraft to search for an exotic Planet-Nine-like object through its
gravitational field~\cite{Witten:2020qew}; the same logic serves a different purpose here.
Once a hidden companion has been dynamically localised, a probe near its gravitational
projection would also reduce the ordinary propagation distance in any brane-to-brane signal
channel. A probe with asymptotic speed \(v_\infty\) covers the distance \(D\) in
\begin{equation}
  t_{\rm travel}
  \simeq
  95\,\mathrm{yr}
  \left(\frac{D}{2000\,\AU}\right)
  \left(\frac{100\,\mathrm{km/s}}{v_\infty}\right),
  \label{eq:travel_time}
\end{equation}
and Table~\ref{tab:travel} gives a few representative speeds.

\begin{table}[h]
\centering
\begin{tabular}{ccc}
\toprule
\(v_\infty\) & Speed [\AU/yr] & Time to \(2000\,\AU\) \\
\midrule
\(17\,\mathrm{km/s}\)  & 3.6  & \(\sim560\,\mathrm{yr}\) \\
\(50\,\mathrm{km/s}\)  & 10.5 & \(\sim190\,\mathrm{yr}\) \\
\(100\,\mathrm{km/s}\) & 21.1 & \(\sim95\,\mathrm{yr}\) \\
\(200\,\mathrm{km/s}\) & 42.2 & \(\sim48\,\mathrm{yr}\) \\
\(300\,\mathrm{km/s}\) & 63.3 & \(\sim32\,\mathrm{yr}\) \\
\bottomrule
\end{tabular}
\caption{Travel time to \(D=2000\,\AU\) as a function of asymptotic probe speed.
A Voyager-like speed of \(17\,\mathrm{km/s}\) is far too slow for practical exploration.
A dedicated fast probe with \(v_\infty\sim100\text{--}300\,\mathrm{km/s}\) reaches the
target region in decades to a century.}
\label{tab:travel}
\end{table}

A Voyager-class speed puts the mission well beyond a human lifetime; only a dedicated fast
probe brings it down to a plausible decades-to-a-century range. Communication with the
probe itself, once there, is comparatively mild: the one-way light travel time to
\(D=2000\,\AU\) is
\begin{equation}
  t_{\rm light}
  \simeq
  11.5\,\mathrm{days}
  \left(\frac{D}{2000\,\AU}\right),
\end{equation}
corresponding to a round-trip delay of order three weeks. That rules out real-time
control, but it is negligible compared with the multi-decade travel time and with an
orbital period measured in tens of thousands of years.

Once the probe is in the neighbourhood, the companion's own gravity becomes a useful
instrument in its own right. At ordinary spatial separation \(b\) from a companion of mass
\(M_X\),
\begin{equation}
  a_X
  =
  \frac{G M_X}{b^2}
  \simeq
  5.7\times10^{-6}\,\mathrm{m/s^2}
  \left(\frac{M_X}{\Mjup}\right)
  \left(\frac{1\,\AU}{b}\right)^2,
  \label{eq:hidden_acceleration}
\end{equation}
which for a Jovian-mass companion at \(b=1\,\AU\) amounts to a velocity drift of
\begin{equation}
  \Delta v
  \sim
  0.5\,\mathrm{m/s}
  \quad \text{per day}
\end{equation}
---easily within reach of high-precision spacecraft tracking. Long before any
brane-to-brane transmission is attempted, the probe could already use this drift to map
the companion's field and pin down its mass, position, and orbit directly.

It is worth checking this operating scale against the companion's own Hill radius. For an
object at heliocentric distance \(D\),
\begin{equation}
  r_H
  \simeq
  D
  \left(
  \frac{M_X}{3\Msun}
  \right)^{1/3}
  \simeq
  137\,\AU
  \left(\frac{D}{2000\,\AU}\right)
  \left(\frac{M_X}{\Mjup}\right)^{1/3}.
  \label{eq:hill_radius_hidden}
\end{equation}
So working at \(b\sim1\text{--}10\,\AU\) around a Jovian-mass companion at
\(D\sim2000\,\AU\) leaves the probe comfortably inside its sphere of gravitational
influence, \(b\ll r_H\). There is, of course, no surface to aim for on our brane; \(b\)
is simply how close the probe gets, in ordinary three-dimensional terms, to the companion's
gravitational projection.

The local-transmission strategy therefore changes only one piece of the problem, but it
changes it by a wide margin. Staying near Earth is technically simpler, but it samples the
ordinary propagation factor at \(r\simeq D\). Sending a probe to the companion is
difficult, but it reduces the source--receiver separation from thousands of AU to
\(b\sim1\text{--}10\,\AU\), yielding amplitude gains of \(10^2\text{--}10^3\) and flux
gains of \(10^4\text{--}10^6\). None of this supplies the high-frequency coherent
excitation mechanism identified as necessary in the previous section, nor a
compact-direction transfer factor \(\mathcal{T}_n(\omega)\), nor a detector response to the
resulting signal. It shows only that, if such a channel exists, its most favourable
operating point is not near Earth but in the immediate vicinity of the hidden companion.


\section{Conclusions}
\label{sec:conclusions}

Two conclusions emerge from this analysis, and they sit side by side rather than in
tension.

The first is negative, but it is not a minor caveat: a smooth halo population simply
cannot supply a Solar-System-scale hidden-brane star or planetary companion. The local
dark matter density is too thin for that, not marginally but by a wide margin. At
\(1000\,\AU\), the smooth-halo median mass scale is only \(\sim10^{-3}\,\Mearth\), while
anything astrophysically interesting requires at least Earth, Neptune, or Jovian mass.
Folding in the usual MACHO and compact-object constraints does not soften this conclusion;
if anything, it sharpens it further.

The second conclusion is more constructive. If the hidden sector has its own
structure---if it can cool, fragment, and bind objects rather than remaining a featureless
collisionless gas---then a real window opens up. Solar-System dynamics, at the level of the
illustrative tidal envelope used here, allows hidden companions of roughly
\(1\text{--}40\,\Mearth\) at \(300\text{--}1000\,\AU\), rising to Saturn or sub-Jovian
masses by \(1350\text{--}1500\,\AU\), and to Jovian mass near \(2000\,\AU\). What makes
this more than a bare mass-distance window is that a simple hidden version of QCD, with a
confinement scale larger than the ordinary one by a factor \(\kappa\sim10\), places the
minimum hidden stellar mass in the upper part of this same range. This gives a concrete
benchmark for a hidden-sector star: bright in its own dark photons, but electromagnetically
dark to us.

Such a companion would not be a smooth-halo MACHO in the usual sense, and the statistical
abundance limits built for halo populations do not directly decide its viability. Finding
or excluding it is a different kind of problem, one that belongs to local Solar-System
phenomenology: high-precision ephemerides, targeted astrometric or lensing searches,
trans-Neptunian dynamics, long-period comets, and, eventually, direct spacecraft tracking.

There is also a more speculative reason to care about exactly where such a companion would
sit. In the companion paper, the KK tower was treated as a possible gravitational
information channel between two sectors localised on different branes and separated by a
microscopic distance in an extra dimension. In that setting, a hidden companion of the Sun
would matter not because it is guaranteed to emit a useful signal, but because it would be
the nearest plausible hidden-sector bound object. It could therefore be the best local
place to look for a natural brane-to-brane gravitational signal, or the best target or
relay for any deliberately generated one. This is the motivation for asking how the
geometric signal scale changes across the allowed mass-distance window.

For a propagating KK mode, the Earth-based received source proxy is proportional to
\((GM/c^2)G_n(\omega,d)\), which reduces to \(GM/(c^2d)\) in the far-field regime
\(G_n(\omega,d)\sim e^{ik_n d}/d\). Along the illustrative ephemeris envelope, where
\(M_{\rm max}^{\rm eph}\propto d^3\), this proxy grows as \(d^2\): counter-intuitively, the
most favourable companion for this purpose is not the nearest one allowed, but the most
massive one still allowed at the largest dynamically relevant distance. Sending a probe
close to the companion's gravitational projection would help for the same reason. It would
replace the ordinary Earth--companion separation \(D\) by a much smaller probe--companion
separation \(b\), yielding amplitude gains of \(10^2\text{--}10^3\) and flux gains of
\(10^4\text{--}10^6\) for \(b\sim1\text{--}10\,\AU\) and \(D\sim2000\,\AU\). None of this
says that such a signal exists, or that it would be strong enough to detect. It only
identifies where, if such a channel were present, the ordinary geometric penalty would be
smallest.

These conclusions should be read with several caveats in mind.

The ephemeris envelope used throughout is illustrative rather than exact. A genuine
constraint would depend on sky direction, orbital phase, eccentricity, inclination, and the
time baseline of the ranging data~\cite{Fienga:2020,HolmanPayne:2016}, none of which enters
the simple scaling \(M_{\rm max}^{\rm eph}(d)=Kd^3\) used here. That envelope is calibrated
against data at \(500\text{--}650\,\AU\), and stretching it out to several thousand AU
should be read as an order-of-magnitude guide, not a precise bound.

A second caveat concerns the companion's very existence. A hidden-brane object at hundreds
or thousands of AU requires a structured hidden sector, and in the stellar or planetary
reading explored here a dissipative one. How such a system would actually have formed,
what its capture probability might have been, what a dark accretion environment would look
like, or whether it might be part of some larger multi-body structure has not been modelled
here.

The source-strength proxy used above is a third limitation, and the most important one. It
was never meant as a detection forecast, and it should not be read as one. Knowing that a
companion has a favourable mass and distance says nothing by itself about whether it would
produce a usable gravitational signal. A genuine prediction would require knowing how
efficiently the hidden source converts its own energy into a coherent, appropriately
structured gravitational disturbance, parametrised here by \(\epsilon_n(\omega)\); how
strongly the relevant KK mode connects the two branes, encoded in
\(\mathcal{T}_n(\omega)\); how that disturbance propagates across ordinary space, described
by \(G_n(\omega,r)\); the actual KK mass spectrum and thresholds; and the response of a
real detector. None of these has been computed here. What can be said is only kinematic:
for the benchmark used in this paper, \(L=0.1\,\mu{\rm m}\), the first KK mode sits at
\[
  f_1\simeq1.5\times10^{15}\,\mathrm{Hz},
  \qquad
  m_1c^2\simeq6\,\mathrm{eV},
\]
far beyond anything ordinary macroscopic motion of a companion---whether orbital or
internal---could ever reach, as Section~\ref{sec:source_strength_proxy} worked out in
detail.

Finally, the scaling relation in Eq.~\eqref{eq:Mmin_kappa} for the minimum hidden stellar
mass is a parametric estimate and nothing more. A serious calculation would have to specify
the hidden nuclear binding energies, some hidden analogue of the proton-proton chain or
another energy-generating reaction, and the opacity as a function of \(\alpha'\),
\(m_{e'}\), and \(\kappa\).

What remains, then, is to turn the allowed companion window into an actual signal
calculation. Concretely, this means evaluating Eq.~\eqref{eq:Sn_general} with real inputs
for the relevant KK modes, the brane wave-function overlaps, the actual source--receiver
separation, and a concrete model of the source dynamics, and then comparing the resulting
signal against a realistic detector response. Only that calculation would settle the
question this paper has only framed: whether a hidden companion of the Sun, if one exists
in the window identified here, could ever act as a detectable source, target, or relay for
a brane-to-brane gravitational channel, or whether it would remain only a dynamically
permitted but silent hidden-sector neighbour.


\section*{Acknowledgements}

I thank Marco Cirelli for useful discussions and helpful comments.

\begingroup

\end{document}